\documentclass{article}
\usepackage{graphicx} % Required for inserting images

\title{Courses}
\author{s.mahdavi.hezavehi }

\usepackage{natbib}
\usepackage{array}
\usepackage{booktabs}

\title {Methodological Considerations for Self-adaptive Systems: An Essay}
\author{Sara Mahdavi Hezavehi \\Supervisors: Danny Weyns and Paris Avgeriou}

\begin{document}

\maketitle

\section{Introduction}

In this essay, we provide an overview of methodological considerations necessary to lay out the foundation for our PhD research on uncertainty and risk-aware adaptation.  

\section{Self-adaptive systems}
Software systems operating in uncertain circumstances require human supervision to adapt to uncertainties. These uncertainties and the consecutive need for adaptation in a software system may originate from a variety of anticipated or unforeseen factors. This includes new stakeholders’ requirements, changes in the availability of resources and requirements priorities, or changes in the environment itself. 

As the complexity of these software systems and the underlying uncertainty (e.g., sources of uncertainty may change, appear or disappear) increase, so does the need for human intervention in order to ensure  continuous and seamless system operation. Over time, managing the adaptation of such systems becomes time-consuming, costly, and requires a great deal of human effort. Therefore, handling uncertainties and at the same time decreasing the amount of human involvement in a running system lead to the need for designing and implementing \textit{self-adaptive systems}. To put it simply, the idea behind designing a self-adaptive software system is to enable the system to modify its runtime behavior in order to deal with a dynamic system context, and changing or new system requirements (in other words uncertainties) in order to function uninterruptedly and provide dependable services while minimizing the need for interventions \cite{deLemos2013},  \cite{weyns2020book}. 

To deal with uncertainty, self-adaptive systems make runtime choices to apply specific adaptation actions. While these adaptation actions (e.g., adding new service providers in a service-based system) mostly focus on realizing one specific goal (e.g., supporting a defined level of service availability in a service-based system), other system goals (e.g., cost of the new service provider) may be negatively affected.

Over time, aggregation of these negative impacts may cause the self-adaptive system to progressively diverge from achieving system goals such as low operational cost and latency, or other quality properties. Therefore, to perform an adaptation that deals with uncertainty, a self-adaptive system is expected to select an \textit{adaptation option} (i.e., a new system configuration) that not only fulfills the systems' functional requirements and adaptation goal (e.g., availability in a service-based system), but also offers advantages such as lower costs and operational time, and potentially sustains the required levels of system-specific quality attributes such as performance, or security. 

Selecting one adaptation option, from the often sizable adaptation or solution space, (i.e., the set of possible adaptation options) that balances these \textit{multiple concerns}, which may also vary or increase in size overtime, can become a substantially challenging task. Therefore, it is essential to equip the system with the right tools and techniques to overcome this challenge. However, due to the high complexity and heterogeneity of self-adaptive systems, designing and developing such systems are intricate tasks.

%To adapt themselves, or more precisely, to make runtime adaptation decisions, self-adaptive systems often follow a set of steps resembling functions of a \textit{feedback loop}. To that end, a self-adaptive system constantly monitors its own and its environment's state, collects relevant information, analyses the collected information to spot any possible abnormalities (e.g., faulty system functions or a safety breach), decides whether or not to adapt the system, plans for adaptation, and applies the selected adaptation option. However, due to the high complexity and heterogeneity of software systems, designing and developing such systems are intricate tasks.

One well-established way to tackle this complexity is to divide self-adaptive systems into two main separate subsystems: 1) a subsystem accommodating the feedback loop to maintain the self-adaptation capabilities (i.e., the managing subsystem), and 2) a subsystem that is responsible for fulfilling users' concerns and ensuring that the system is satisfying its original goals (i.e., the managed subsystem) \cite{weyns2020book}.
In the context of this PhD thesis, the self-adaptive systems under investigation, respect this logical division of sub-systems.\\

\subsection{Main concepts} 

A \textbf{self-adaptive system} (SAS) is  \textit{a system which, in face of uncertainty, is capable of autonomously modifying its runtime behavior in order to achieve system objectives}. To achieve that, a self-adaptive system continuously monitors itself and its environment, gathers relevant data and analyzes them to decide if adaption is required.

%Design and implementation of such software systems can be a challenging task. A self-adaptive system not only must be able to apply changes at runtime to meet the functional requirements of the system, but similar to any other software system, needs to fulfill other system concerns such as performance and availability up to a satisfying level. 

Central to the problem of handling uncertainty at runtime, is the challenge of dealing with multiple system concerns deemed essential for the stakeholders. 
\textbf{Multiple system concerns} in the context of this thesis refer to: a) the adaptation benefits obtained in terms of  {quality requirements \footnote{Also referred to as non-functional requirements, quality properties, quality attributes.}; b) the cost of adaptation; and c) the potential risks (e.g., safety or privacy risks) that may arise due to the application of adaptation, are relevant to the stakeholders of the system and need to be handled to ensure public acceptance of the system as well.} 
%It is worth noting that in the initial phases of this PhD research the focus of our efforts was only on \textit{multiple quality attributes}. However, based on the knowledge we gained by conducting the first and second studies (reported in chapters \ref{chap:2} and \ref{chap:3}), we broaden our focus to include other important factors which should be taken into account during runtime decision-making, hence, redirecting our efforts toward \textit{multiple system concerns} instead of \textit{multiple quality attributes}.    

%\textit{Handling uncertainty in the presence of multiple concerns} implies that not only the system should be able to correctly identify and locate the situation triggering the need for adaptation (e.g., uncertainty), but also prioritize the adaptation options to choose the alternative which meets the concerns at a satisfactory, if not optimal, level and execute adaptation while  handling implications of adaptation (e.g., lower performance or latency). Naturally, in the presence of multiple concerns, and a sizable solution space for adaptation options, the complexity of the decision-making process for making adaptation decisions to handle uncertainty increases.

To address those multiple system concerns, a self-adaptive systems is equipped with mechanisms called feedback loops, also known as control loops.
\textbf{Feedback loops} are commonly used to \textit{provide the system with the ability to continuously monitor itself, collect data about the designated system components in the managed subsystem, perform analysis, make decisions, and follow suitable steps to implement and execute required adaptation actions.} Self-adaptive systems typically use one feedback loop to enable self-adaptation capabilities, however, some systems may be designed to take advantage of two or multiple feedback loops which interact with each other \cite{Tichy, Nakagawa}.

The \textbf{decision-making} module of a self-adaptive system is \textit{the module responsible for deciding which adaptation option should be selected and applied}. This module is of paramount importance for successful adaptation at runtime. By designing a module that takes relevant system concerns into account and performs efficient trade-off analysis, the outcome of the decision-making mechanism can be enhanced to select the best adaptation option from the solution space. It is therefore essential for software architects and designers to envisage and propose design-time solutions (e.g., architecture tactics, simple equations, etc.) that can be translated, implemented, and used by the decision-making mechanism at run-time. \\

\section{Architecture-based self-adaptation}
The architecture of a software-intensive system consists of design decisions and is considered to be a blueprint of the system construction and evolution \cite{Jansen}. Such design decisions cover all aspects of the system under development including structure of the system, system components, interactions between the components, their behavior, and specifications of quality properties. \\
The software engineering community had initially focused predominantly on static (i.e., non-adaptive) software architectures supporting architects in their decision making process \cite{Kruchten, Jansen, Bass}. With the advent of complex self-adaptive software systems, the use of software architecture as a solution for designing dynamic software systems led to a gradual emergence of many architecture-based adaptation approaches in the last two decades. \\
In recent years, architecture-based approaches have been widely used to design and build adaptive software systems. In these approaches, the system is equipped with a feedback loop maintaining \textit{a model of itself} to continuously observe its own state and evaluate the model for requirements compliance. If needed, the system can apply adaptations either at system, module, or parameter level \cite{WeynsD}.\\
One of the core principles of software architecture is \textit{separation of concerns} \cite{Oreizy1999}, \cite{Andersson2009}. This principle manages the complexity of system development through partitioning of the software system in such a way that each module is responsible for a particular concern. The aim is to minimize the overlap between these modules, which in addition to complexity management, facilitates independent maintenance and evolution of each module over time. \\
Architecture-based adaptation approaches follow the principle of separation of concerns by distinguishing between the domain concerns that are handled by the \textit{managed subsystem} (i.e., what the system should provide to the user in terms of functions or services), and adaptation concerns that are handled by the \textit{managing subsystem} (i.e., how the domain concerns are achieved in terms of benefit, cost, and risk). This characteristic of architecture-based solutions grants enough flexibility to modify the system both at design- and runtime to support the dynamicity of uncertainties that may appear or disappear due to changes and affect the system in unpredictable ways.\\
In this PhD project, we investigate and contribute to the field of architecture-based self-adaptation in order to support uncertainty handling in the presence of multiple concerns throughout the system lifetime, from designtime to runtime.

\subsection{MAPE-K reference model}
The essential functions of architecture-based self-adaptation can be supported by the well-known MAPE-K (i.e., Monitor, Analyze, Plan, Execute, and Knowledge component) reference model. MAPE-K was first introduced by \cite{ibm2005architectural} in the domain of autonomous systems and later was adopted and widely used in the design of self-adaptive systems. 
By complying with the concept of separation of concerns (i.e. separation of domain-specific concerns from adaptation concerns), the MAPE-K model also supports reusability and manages the complexity of constructing self-adaptive systems.

The MAPE-K reference model consists of two major sub-systems: 1. a managing sub-system and, 2. a managed sub-system. The managing sub-system comprises the adaptation logic that deals with one or more concerns by continuously looking into the managed sub-system; this happens through its five main functionalities (i.e., Monitor, Analyse, Plan, Execute, and a shared Knowledge component). The managed sub-system comprises the application logic that provides the system’s domain functionality.
The MAPE functionalities map to the basic functions of a feedback loop, while the K component maps to runtime models maintained by the managing system to support the MAPE functions \cite{journals/taas/WeynsMA12}.

In the context of this PhD thesis, the MAPE-K model is often used as reference to communicate and expose the architectural aspects (e.g., system elements and their relations) of self-adaptive systems when required.

\section{Research design}\label{sec:intro:research-design}
\subsection{Problem statement}\label{sec:intro:problem-statement}
The \textit{concept of uncertainty}, its sources and characteristics have long been the focus of many research endeavours \cite{Survey2021, Keith2019, Esfahani2013, Ramirez2012, Garlan2010} in the self-adaptive systems domain. Although existing approaches and findings on uncertainty resulted from these research efforts (e.g., uncertainty matrix, uncertainty taxonomy, and terminology as well as a topology of uncertainty) are valuable and potentially reusable in the domain of self-adaptive systems, they suffer from three major shortcomings: 
\begin{enumerate}
  \item The efforts to unravel the concept of uncertainty are often not conducted systematically and incrementally to the previous work, hence, the connection between various existing taxonomies, definitions and terminologies is rather unclear. This, at times, resulted in isolated research efforts which are hard to be incorporated or applied in the realm of self-adaptive systems.
    \item Most existing work is problem-specific (e.g., model-based decision support presented in \cite{Walker}, water management modelling process  of \cite{REFSGAARD20071543}, patterns for Self-Adaptation in Cyber-Physical Systems \cite{Musil2017}) and therefore difficult  to reuse and re-apply in different disciplines of the self-adaptive system domain.
    \item Current approaches have been proposed top-down and have never taken the perspective of the research community into account. Although uncertainty is increasingly being considered as a first-class citizen in the self-adaptive systems domain, and selection of methods to deal with uncertainty is more and more tailored-based depending on the specific type and source of uncertainty \cite{Moreno:2015, Cmara2017UncertaintyIS, 10.1145/3194133.3194145, 10.1145/3328730}, the community's collective perception on what constitutes uncertainty has never been compiled, synthesized, and taken into account when developing uncertainty handling methods. This lack of a unified view and knowledge on what key characteristics of uncertainty must be considered from the perspective of the research community is an essential factor hindering proposing effective approaches which can be valuable in practice. 
    \end{enumerate}

In this PhD research, we focused on the following problem statement to address these shortcomings:\\

\textit{What is the nature of uncertainty in self-adaptive systems and how can this uncertainty be managed in a systematic, incremental and domain-independent way to address stakeholders' concerns?}\\

To study the nature of uncertainty, we need to obtain a better understanding of how uncertainty is currently perceived in the literature and pinpoint the areas where the link between existing work requires improvement. The collective knowledge gained from studying the existing tailored-based methods for handling uncertainty can lead us to establish a basic set of best practices in an effort to resolve the limitations of current methods. 

%This, in turn, prompted us to resolve the shortcomings by proposing solutions with a coherent outlook on the concept of uncertainty in the domain. %

%Equipped with a unified view of uncertainty in the domain of self-adaptive systems,% 
Next, we will investigate the concept of uncertainty from the perspective of the research community, expressed through various stakeholders, to identify their specific needs. We aim to put forward a method for managing uncertainty which is efficient to reuse while addressing the common concerns of stakeholders effectively.

In the context of this PhD research, stakeholders' concerns  may refer to a variety of issues such as adaptation benefits obtained in terms of quality requirements, the monetary aspect of adaptation, or domain specific risks such as safety or breach of privacy that may arise due to the application of adaptation.

\begin{figure}[]
    \centering
    \includegraphics[width=\textwidth]{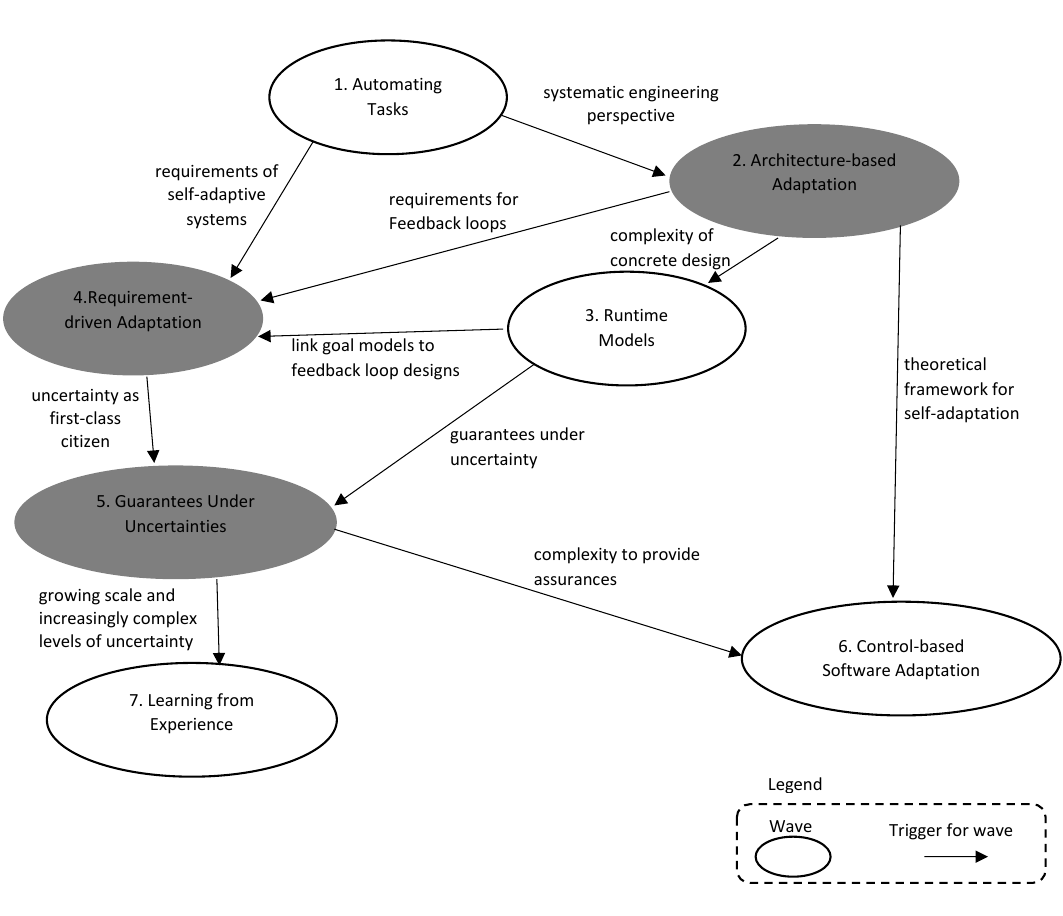}
    \caption{The seven waves of research on engineering self-adaptive systems \cite{weyns2020book}.}\label{fig:intro:wave}
\end{figure}

Figure \ref{fig:intro:wave} shows the \textit{seven waves of research on engineering self-adaptive systems} that have emerged over time \cite{weyns2020book}. Each wave is indicated using an oval and are numbered chronologically based on the time it has been appeared. The arrows show how insights from one wave initiated the new waves \cite{weyns2020book}. 
To tackle this problem, this PhD research contributes to three waves indicated in grey ovals in Figure \ref{fig:intro:wave}.

To summarize, we first shed light on the concept of uncertainty to address the general lack of consensus on what uncertainty actually entails in the domain of self-adaptive systems (i.e., Guarantees under Uncertainty wave). Next, we studied the existing approaches tackling uncertainty in self-adaptive systems and proposed an architecture-based method (i.e., Architecture-based adaptation wave), that in addition to quality concerns of the system, accounts for the major stakeholders' concerns, namely costs and risks of adaptation as pivotal factors during runtime decision-making (i.e., Requirement-driven Adaptation wave).

%Our efforts to tackle the problem posed in this section initially helped to get an overview of the existing architecture-based approaches for handling uncertainty in self-adaptive systems, and then contributed to identification of concerns in architecture-based self-adaptive systems, clarification of uncertainty notion in architecture-based self-adaptive systems, and led us to proposing architecture-based solutions to deal with uncertainty in self-adaptive system both in designtime and runtime phases.  

\begin{quote}\itshape

\end{quote}

\subsection{Design science as research methodology} \label{Design science}
This PhD research project adopts the design science framework proposed by  \cite{Wieringa2014}. %Design science is a problem-solving paradigm aiming at use of innovative ways to define ideas, practices, technical capabilities, and products through which the analysis, design, implementation, and use of information systems can be effectively and efﬁciently achieved \cite{Hevner}.
In principle, design science concerns the design and investigation of artefacts in context. Design refers to allowing the design of an artefact that improves a problem context, and investigation refers to allowing to answer knowledge questions about the artefact in the context \cite{Wieringa2014}. Design problems - also known as technical or practical problems- call for a change in the real world, while knowledge questions ask for knowledge about the real world. 

\begin{figure}[]
    \centering
    \includegraphics[width=0.8\textwidth]{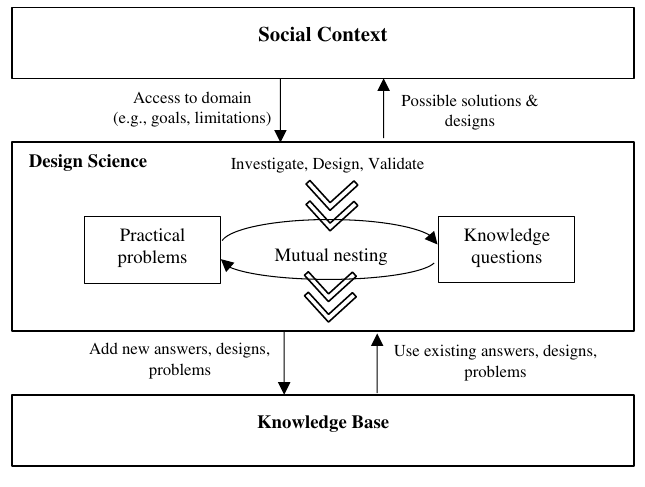}
    \caption{A simplified overview of the design science framework by Wieringa.}\label{fig:design science}
\end{figure}

Figure \ref{fig:design science} shows a simplified overview of the design science framework as proposed by Wieringa. In design science, practical problems and knowledge questions are mutually nested; this means that design science allows creating nested problem solving cycles, in which solving a practical problem may lead the problem solver to ask knowledge questions, and answering these knowledge questions may lead the problem solver to new practical problems \cite{WieringaRoel}. But it is important to note that practical problems and knowledge questions' require different approaches to address them, and therefore, they should be distinguished from one another. Common strategies to tackle knowledge questions includes empirical cycles (i.e. performing empirical studies) while solving practical problems entails design cycles (i.e., problem investigation, treatment design and validation).\\
Design science as a conceptual framework is an iterative process where the problem solver analyses the problem, proposes a solution, evaluates the solution, and if needed, iterates from the beginning. This "cyclic" nature of design science helps the problem solver to identify and better understand different aspects of the design problem, which may result in posing new  practical problems or knowledge questions leading the problem solver to find satisfactory solutions. \\
These characteristics of design science makes it a great candidate to be applied in the context of a PhD research project where the researcher needs to address a design problem by decomposing it to practical problems or knowledge questions, acquire knowledge, and identify solutions for the posed nested problems, thus eventually resolving the original design problem. \\
Following the guidelines of Wieringa, we created a detailed nested problem decomposition consisting of several knowledge questions and practical problems as elaborated in Section \ref{c1:sec:problem-decomposition}.

\subsection{Problem decomposition}\label{c1:sec:problem-decomposition}

\begin{figure}[]
    \centering
    \includegraphics[width=\textwidth]{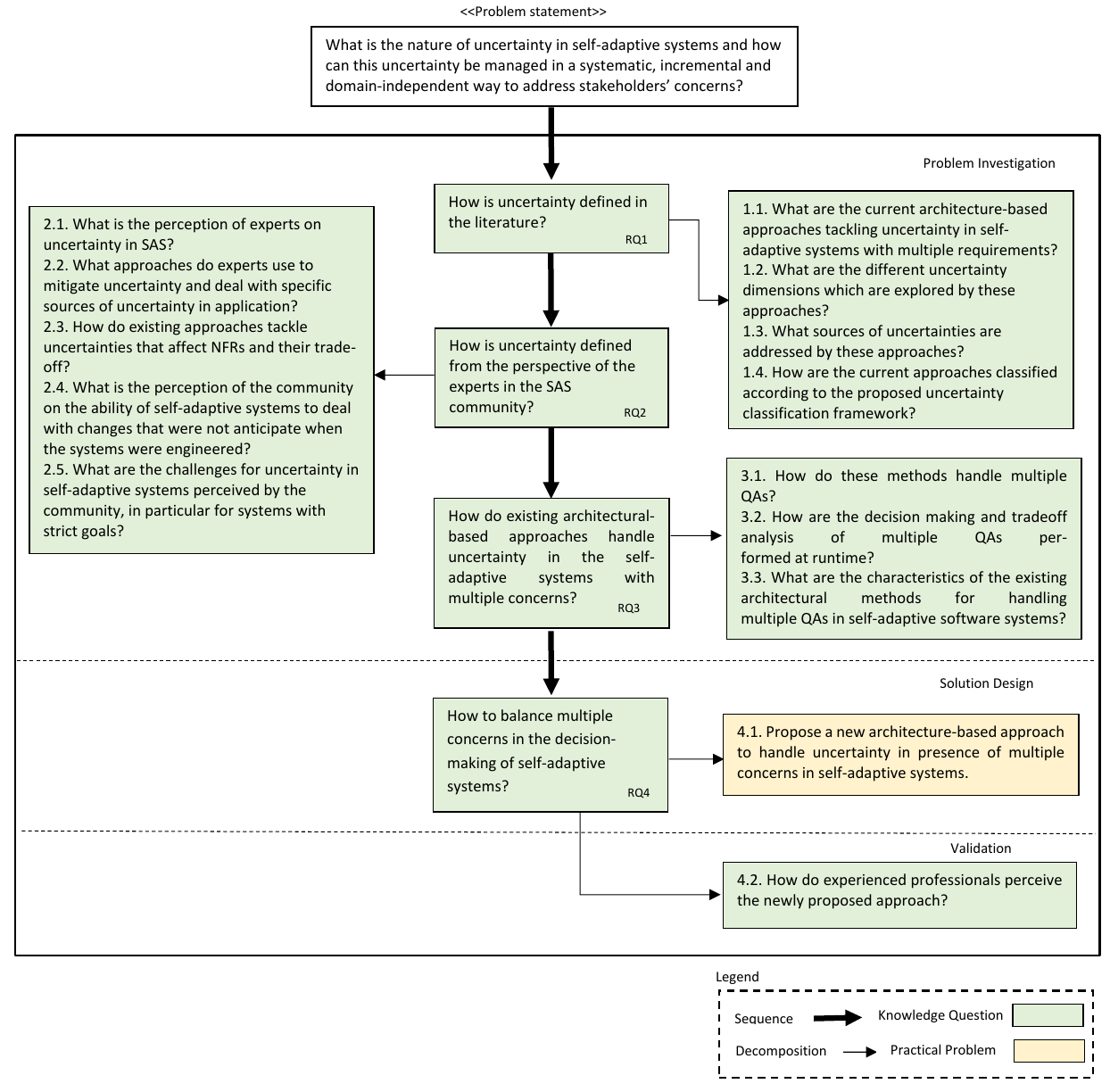}
    \caption{The decomposition of the design problem tackled in this dissertation.}\label{fig:intro:problem-decomposition}
\end{figure}

Figure \ref{fig:intro:problem-decomposition} shows an overview of the nested problem decomposition in this PhD project. In this figure, the main problem statement is addressed using four main research questions (marked with RQ1 to RQ4), which are decomposed into a series of knowledge questions indicated in green boxes and one practical problem indicated using a yellow box. The numbers indicate the logical flow of the questions, and the arrows indicate sequence or decomposition. In this research project, the knowledge questions and the practical problem were addressed in three different stages: problem investigation, solution design, and validation.

The first stage of the problem decomposition focuses on problem investigation. We started this stage by posing RQ1 (i.e., "How is uncertainty defined in the literature?") regarding the concept of uncertainty in the state-of-the-art. We decomposed this question into four sub-questions which we addressed in a systematic literature review. This comprehensive systematic literature review explores existing architecture-based approaches tackling uncertainty, the notion of uncertainty itself, its characteristics, different types, and sources of uncertainty in the domain of self-adaptive systems. While conducting this review, we noted that in addition to a lack of consensus on what constitutes uncertainty, structured ways of defining uncertainty and its characteristics in the self-adaptive system domain are not accounted for as well. This work is reported in chapter 2 of this thesis.

The conducted literature review (i.e., answer to RQ1) revealed varied, vague and at times non-conforming definitions for uncertainty in the existing work. Therefore, we decided to investigate uncertainty from the perspective of the domain experts to learn their take on the concept of uncertainty and how they perceive uncertainty despite the inconsistencies in the literature. To that end, we posed a new research question (i.e, "How is uncertainty defined from the perspective of  the experts in the SAS community? "). In addition to probing experts' perception on uncertainty, we investigated how they manage the impact of uncertainty on non-functional requirements (i.e., quality attributes of the system), we studied their remedies for unanticipated changes, as well as, challenges they face while dealing with uncertainty. The insights obtained from answering this research question were meant to help us to come up with solutions that could potentially better fulfill practitioners’ needs. This work is reported in chapter 3 of this thesis.

To complement insights we gained from answering research question two, which focused on experts' perceptions on uncertainty and methods to deal with it in practice, we proposed research question three (i.e., "How do existing architectural-based approaches handle uncertainty in the self-adaptive systems with multiple concerns?") to study characteristics of existing architecture-based methods for handling uncertainty in the literature. In an effort to identify and shed a light on stakeholders' concerns as defined in the literature, and to spot possible discrepancies or similarities with experts' concerns in practice (i.e., answers to RQ2) derived previously, here we predominantly focused on how quality attributes of the system and trade-off analysis are addressed. This work is reported in chapter 4 of this thesis.
%Amongst others, results suggested that major concerns in a self-adaptive system can be categorized under the umbrella terms "cost" "benefit", and "risk".  

The outcome of the initial stage (Problem Investigation), which included both experts' input as well as literature analysis, indicated that further work is required to improve existing methods both in terms of addressing uncertainty as a concept, as well as, stakeholders' concerns while applying adaptation to handle uncertainty. %In particular, new methods may be proposed  to refine trade-off analysis of multiple concerns and enhance the decision making mechanism at runtime.
This led us to propose the fourth research question (i.e., "How to balance multiple concerns in the decision-making of self-adaptive
systems?"). %which led us to a practical problem focusing on the development of new architecture-based approaches. 

Aiming to propose reusable means for expressing, modeling and handling  concerns, we decomposed RQ4 into two additional questions (i.e., PP 4.1, and KQ 4.2). To answer practical problem 4.1 (i.e., "Propose a new architecture-based approach to handle uncertainty in presence of multiple concerns in self-adaptive systems."), we put forward an architectural viewpoint which considers the estimated benefit in terms of quality attributes of the system, adaptation costs, and risks as core factors during decision-making of adaptation. The proposed approach aims at helping software architects to define appropriate designtime strategies and create generic templates or solutions for data collection and decision-making mechanism which then can be instantiated at runtime.

In Validation stage, to demonstrate the applicability, usefulness, and understandability of the proposed viewpoint, we addressed knowledge question 4.2 (i.e., "How do experienced professionals perceive the newly proposed approach?"). To answer this knowledge question, we designed and conducted a case study in which participants with experience in the engineering of self-adaptive systems performed a set of design tasks in an Internet-of-Things (IoT) system named DeltaIoT. As part of the case study, we conducted a survey to collect data in order to: 1) cross-validate the retrieved data with the findings of the case study where applicable, and 2) get feedback to determine whether or not we have covered their main concerns as stakeholders and to further improve the viewpoint. The results of the case study confirmed an overall positive outlook of participants on the viewpoint. This work is reported in chapter 5 of this thesis.

\subsection{Mapping empirical methods to the RQs}
As discussed in Section \ref{Design science}, different knowledge questions and practical problems may require a variety of research methods to be addressed and evaluated. In Table \ref{c1:tab:overview-methodology}, we list the research methods we used to tackle the questions posed in the problem decomposition of this PhD research as indicated in  Section \ref{c1:sec:problem-decomposition}.
\begin{table}[]
   \footnotesize
    \centering
    \caption{Empirical methods used to answer the knowledge questions.}
    \label{c1:tab:overview-methodology}
    \begin{tabular}{@{}lm{3.75cm}m{2.75cm}m{2cm}l@{}}
    \toprule
    \textbf{ID} & \textbf{Research question} & \textbf{Empirical}\newline \textbf{method} & \textbf{Data}\newline\\ \midrule
    RQ1 & \textit{How is uncertainty defined in the literature?} & Systematic literature review & Quantitative \\
    RQ2 & \textit{How is uncertainty defined from the perspective of the experts in the SAS community ?} &  Survey &  Qualitative \& Quantitative   \\
    RQ3 & \textit{How do existing architectural-based approaches handle uncertainty in the self- adaptive systems with multiple concerns?}& Systematic literature review & Quantitative\\
    RQ4 & \textit{How to balance multiple concerns in the decision-making of self-adaptive systems?} & Case study & Qualitative \\
    \bottomrule
    \end{tabular}
\end{table}
Specifically, in this PhD project we adopted three empirical research methods: 1. systematic literature review, 2. survey, and 3. case study. In the following, we briefly describe each of these research mehtods:

\begin{description}
    \item[Systematic literature review] A systematic literature review is a method used to identify, evaluate and interpret available research relevant to a particular research question, topic, or phenomenon of interest \cite{Kitchenham}. A systematic review helps to: 1. summarise the existing evidence with respect to a topic; 2. identify gaps and areas for improvements in the existing research; and consequently, 3. correctly position new research endeavours for future work. The first step in conducting a systematic literature review is creating a protocol, in which all the steps of the systematic review should be defined in detail. More specifically, the protocol contains the research questions, search strategy to identify and collect relevant primary studies, inclusion and exclusion criteria for filtering out irrelevant papers, and methods for extracting data and synthesizing them to answer the  research questions. A systematic review is considered a secondary study, while the individual studies investigated as part of the review are considered primary studies. 
    
    \item[Survey] is a data collection and  measurement process with the aim  of producing statistics in quantitative format \cite{fowler2008survey} or obtain qualitative data  with respect to some aspects of the study population. Collecting data is often done through interviews or questionnaires from a fraction of population (i.e., sample) which is selected based on specific criteria (e.g., convenience). Survey as a research method includes data collection, and data analysis and results in answering the original research questions.

    \item[Case study] can be defined as an empirical method aimed
at investigating contemporary phenomena in their context with an emphasis on using multiple sources of evidence \cite{Runeson2008GuidelinesFC}. Case study methodology can be used for both exploratory and descriptive purposes. In this PhD research, to enhance results obtained from the case study, we followed a Concurrent Triangulation Strategy to validate our collected data \cite{fowler2008survey}. This strategy uses different methods concurrently in an attempt to confirm, cross-validate or corroborate findings. Following the same strategy, we first conducted a case study to collect data in order to evaluate the phenomenon of interest, and then conducted a survey to collect data in order to cross-validate the retrieved data with the findings of the case study, where applicable.
    
\end{description}

\bibliographystyle{apalike}
\bibliography{bibliography}

\end{document}